# Probing the Electrical Switching of a Memristive Optical Antenna by STEM EELS

*David T. Schoen, Aaron Holsteen, and Mark L. Brongersma\**

Geballe Laboratory for Advanced Materials, Stanford University, Stanford, CA 94305

\* Correspondence should be addressed to: brongersma@stanford.edu

**Abstract**

The scaling of active photonic devices to deep-submicron length-scales has been hampered by the fundamental diffraction limit and the absence of materials with sufficiently strong electro-optic effects. Here, we demonstrate a solid state electro-optical switching mechanism that can operate in the visible spectral range with an unparalleled active volume of less than (5 nm)$^3$ or ~$10^{-6}$ $\lambda^3$, comparable to the size of the smallest electronic components. The switching mechanism relies on electrochemically displacing metal atoms inside the nanometer-scale gap to electrically connect two crossed metallic wires forming a crosspoint junction. Such junctions afford extreme light concentration and display singular optical behavior upon formation of a conductive channel. We illustrate how this effect can be used to actively tune the resonances of plasmonic antennas. The tuning mechanism is analyzed using a combination of electrical and optical measurements as well as electron energy loss (EELS) in a scanning transmission electron microscope (STEM).

The tremendous benefit derived from scaling device technologies has served as a central driver of science and technology as well as economic growth for many decades. In electronics, nanoscale components are already commonplace and even single-molecule and single-atom devices are being explored.[1–3] In sharp contrast, the miniaturization of optical components to nanometer-sized dimensions has been faced with a seemingly unsurmountable barrier: the optical diffraction limit. Whereas plasmonics[4] and high-index semiconductor nanostructures[5] have opened up new ways to manipulate light at the nanoscale, downscaling of active elements to subwavelength footprints remains challenging as one cannot benefit from long interaction lengths to achieve a desired functionality. This is particularly true for a range of technology relevant electro-optical switching, modulation[6] and cavity-tuning devices[7–9], which tend to rely on weak electro-absorption or electro-refraction effects[10]. To address this challenge, researchers are actively exploring a wealth of new materials that display very large changes in their optical properties in response to external stimuli, including tailored quantum-wells[11], phase transformation materials[12,13], organics[14,15], 2D conductors such as graphene[16–19], and epsilon-near-zero materials[20–22]. New device physics concepts employing plasmonics[23–25] or optically resonant structures[26] are also being investigated. This general approach to scale active components has been quite successful for the recent development of electrically-driven nanolasers and spasers[27,28]. However, despite many efforts the smallest, power-efficient electro-optical devices have remained micron-scale or larger[29,30]. One key reason for this is that all of the traditional switching/tuning strategies fundamentally rely on the interaction of light with mobile charges (electrons or holes) and electrical gating only affords relatively small changes in their volume density ($\sim 10^{21}$ cm$^{-3}$) before electric fields reach unsustainable levels and electrical breakdown occurs. In this work, we capitalize on the massive changes in mobile carrier density ($\sim 10^{23}$ cm$^{-3}$) that can be achieved without a net charge build up by physically moving metal atoms via an electrochemical process. In our switch-design, these atoms are moved in and out of the nanometer-scale gap between two crossed metallic wires forming a crosspoint junction. As the illumination of such the junctions results in effective light concentration inside the gap via the excitation of surface plasmons, its optical properties are very sensitive to any changes in materials properties in the gap. This then facilitates the realization of a new type of solid state electro-optical switching mechanism that can operate in the visible and near-infrared spectral range with an active volume less than $(5 \text{ nm})^3$ or about $10^{-6} \lambda^3$, i.e. comparable to the size of smallest active electronic components.

Crosspoint junctions are among the most compact and densely integrated electronic devices at $10^{11}$ devices per square centimeter[31]. This architecture has already seen widespread use in molecular electronics, nonvolatile memory[32,33], and most recently memristive logic elements have been used for computation[34–36]. Most of these devices rely on the electric-field induced transport of ionic species to reversibly grow and dissolve a nanoscale conductive filament. This filament has recently been observed to be just a few nanometers in size[37] and its tip is scalable to the single atom level[38]. Because the total volume of this filament is extremely small, the charge transport required to form it is also very small. This facilitates high switching speeds (<1 ns or >1 GHz) and very low switching energy (< 100 fJ)[39]. These desirable features are combined with a high retention time (years) and impressive endurance (>$10^{12}$ cycles). Whereas these desirable traits already opened up a wide range of electronics applications, there have not yet been reported applications in photonics.

**Basic Electrical and Optical Properties of a Memristive Optical Antenna**

Figure 1A shows a schematic of a metallic crosspoint junction designed to elicit a significant change in an optical scattering response by moving about $10^2 - 10^3$ metal atoms in and out of the junction region. The design is inspired by a conventional electrochemical memory device in which the active electrode is a chemically synthesized silver (Ag) nanowire, the ion conducting layer is a 5-nm-thick insulating aluminum oxide ($Al_2O_3$) layer grown by atomic layer deposition (ALD as shown in Fig. S1), and the inert counter electrode is an evaporated gold stripe defined by electron beam lithography (EBL)[35,40]. The application of a positive bias is known to cause oxidation of the Ag electrode, which drives out Ag cations. The cations subsequently migrate to the inert counter electrode where they are reduced to form a metal Ag filament that grows back towards the Ag electrode. Previous research on Au-$Al_2O_3$-Ag junctions suggested that metallic silver filaments of just a few nanometers in diameter can form in this materials system[41]. The inset to Fig. 1B shows a scanning electron microscopy (SEM) image of a tested crosspoint junction and a zoom-in of this junction. To carefully study the switching event, we induced filament growth by sourcing a current through the junction and ramping up the current logarithmically from the pA level to 5 µA over several seconds. When electrical breakdown was achieved, a sudden kink in the IV can be observed. For the fabricated junctions, this typically occurred at a threshold voltage around 3 V (Fig. 1B). After the initial breakdown event takes place, the IV characteristics vary stochastically. This behavior is often referred to in nonvolatile memory studies as the 'forming' stage[36], in which the filament is created through Ag ion migration. After the filament is formed, the device follows a different return trajectory with a much lower resistance. In this particular device the DC resistance dropped by more than three orders of magnitude. The DC resistance is checked by a voltage-sourced IV scan with a small current compliance of 1 nA to avoid making changes to the junction. The measured resistance before breakdown is greater than 1 GΩ, and after breakdown is around $6 \times 10^2$ kΩ. This resistance is still substantially larger than that of a single-atomic point contact $R_0 = (2e^2/h)^{-1} \approx 12.9$ kΩ.[42] This is consistent with previous work which indicates the formation of a filament can abruptly be terminated when using a small current compliance during the growth phase. This is attributed to the rapidly decreasing electronic tunneling resistance from the filament to the electrode with decreasing gap-size that switches off the ionic current flow and thus the filament growth. This results in a final, switched device state for which the total resistance is composed of a larger tunnel resistance in series with a smaller filament resistance.[43,44]

In general, the optical detection of a metallic filament that is just a few nanometers in size would be extremely challenging. However, unpolarized white-light scattering images taken from several devices show a clear change in the light scattering response upon dielectric breakdown in the junction. Figures 1C-D show an example where the region near the termination of the Ag nanowire changes in color from blue to green. This change occurs without an observable modification in the junction geometry in the optical or SEM images. The spectral changes were confirmed and quantified by taking confocal, white-light scattering spectra collected from this area. The overall scattering spectrum shows a noticeable redshift upon switching the device, suggesting that the change in the conductive state of the junction plays an important role in the resonant scattering behavior of the antenna (Fig. 1E).

In analyzing the structure of the multi-peaked scattering spectrum before switching (blue curve), it is clear that several optical resonances must be contributing to the scattered light signal in the far-field. One can clearly identify the presence of the localized surface plasmon resonance of the Ag nanowire peaked at 480 nm[45]. In addition, one would expect to see a series of Fabry-Pérot (FP) type resonances for surface plasmon polaritons (SPPs) that can propagate back and forth between the Ag nanowire truncation and the cross-point junction[46–49]. When the length of this nanowire section is taken to be $L$, one would expect resonances to occur whenever the SPP roundtrip phase $\phi_{RT} = 2k_{SPP}L + \phi_{CR} + \phi_{TR}$ equals an integer $m$ times $2\pi$. Here, $k_{SPP}$ is the SPP propagation constant and $\phi_{CR}$ and $\phi_{TR}$ are the reflection phases that the SPP incurs in reflecting from the nanowire crossing point and the nanowire truncation respectively[47]. The value of $m$ is the mode number and indicates the number of anti-nodes in the oscillating SPP current. Using the experimentally derived SPP dispersion relation and reflection phases for this type and size of Ag nanowire from a previous work[49], we calculated the spectral locations where low-order Fabry-Pérot type resonances are expected (green columns in Fig.1E). Upon switching the device, one could expect all of the FP resonances to redshift by several tens of nanometers as the electric connection causes an approximately $\pi/4$ increase in $\phi_{CR}$, as determined from a previous work in which two metallic wires were mechanically placed in electrical contact[45,49]. This is consistent with the overall redshift of the spectrum.

As the SPP resonances are longitudinal resonances, one would expect the maximum impact of these resonances on the scattering spectra collected for transverse magnetic (TM) polarized light for which the electric field is oriented along the length of the Ag nanowire. Furthermore, the lowest-order resonances occurring at the longer wavelengths are expected to be strongest due to the longer SPP propagation length at longer wavelengths[48,49]. The inset to Fig. 1E shows the TM spectra near the expected location of the m = 3 and m = 4 resonance. Indeed two red-shifting peaks are seen in this spectral range. The observed changes in the far-field light scattering measurements provide direct evidence for the significant modification in the scattering properties. The observed changes also appear to be in reasonable qualitative agreement with the changes expected from a basic FP-resonance model. However, the presence of a plurality of overlapping far-field scattering resonances makes it a very challenging (if not impossible) task to construct and verify an exact quantitative model for the scattering spectrum. This is further complicated by the fact that the excitation and collection efficiencies for the different resonances are quite sensitive to the excitation conditions and exact device geometry.

**Characterization of the electrical switching of a crosspoint junction with STEM EELS**

To get around the challenges associated with analyzing the far-field light scattering spectra, we use a scanning electron microscope (STEM) capable of creating high-spatial-resolution maps of both the morphology of the device structure as well as the local density of optical states (LDOS) by performing electron energy loss spectroscopy (EELS)[48–52]. This requires that the device structures be fabricated on 15-nm-thick, free-standing electron-transparent silicon nitride membranes, here obtained from Ted Pella (See Figure S2 and S3). This procedure facilitates an effective visualization of the optical modes supported by the crosspoint devices. To our knowledge this work also constitutes the first EELS study on an active device in which spectral and spatial modifications of the optical properties (LDOS) were followed to analyze its operation.

To investigate the impact of the filament growth on $\phi_{CR}$, we first analyze a crosspoint junction device with an Ag nanowire that is quite long compared to the SPP decay length. Figure 2A shows a STEM image of the junction and Fig. 2B shows a spatial map of the EELS signal before switching for an energy of 1.24 eV, corresponding to a free space wavelength of 1000 nm. Such maps provide direct information on the spatial distribution of the LDOS as the loss probability for electrons passing by the antenna is exactly proportional to the LDOS projected along the electron beam trajectory[52]. A series of nodes and antinodes can be seen close to the junction indicating a periodic variation in the LDOS at a spatial scale that is equal to half the surface plasmon wavelength. The variations are reminiscent of Friedel oscillations in the local electronic density of states near the surface of a metal object[53].

To further analyze this data, we integrated the EELS signal perpendicular to the nanowire axis to generate a line-scan. This line-scan and those for other energy loss values are shown in red in Fig.2C. Previously, it was shown how such line scans can be analyzed to extract the real and imaginary parts of the SPP dispersion as well as the reflection phases for SPPs reflecting from cavity truncations[49]. When we extract the magnitude of $\phi_{CR}$ from linescans at different energy loss values, we can see that large values close to -$\pi$ are achieved, especially for the short wavelengths for which the SPPs are most confined. The red line-scans in Fig. 2C represent the data after the electrical state of the device is changed from a high to a low resistance state. A significant shift in the antinode position relative to the junction location at x = 0 can be observed. When we extract the phase pickup on reflection from this data, we can see that a nearly constant shift of $\pi/4$ is realized across the observed wavelength range.

At first sight, the $\pi/4$ shift in the reflection phase may appear to be unexpectedly large for the nanoscopic change in the geometry. However, changes of this magnitude can be understood in light of a recent observation that nearly-touching metallic nanostructures feature a singular optical response in the limit they become (electrically) touching[54]. At this point, the flow of charge induced by a plasmonic excitation in the junction can dramatically change as transfer of net, mobile charge between the electrodes becomes possible. This in turn can result in easily observable changes in the optical response and the supported plasmonic modes[49,54]. These large changes in optical response can also be understood from circuit models that describe the optical behavior of two closely spaced metal objects in which a load is placed in the gap. Within the context of such a model, the impedance of the load $Z_{load}$ can be treated as a parallel circuit composed of an inductive impedance $Z_f$ formed by the metallic filament and a capacitive impedance $Z_{gap}$ formed by gap between the two electrodes spaced by $Al_2O_3$. Significant changes in the response are expected when the filament grows and $Z_f$ approaches $Z_{gap}$[55–57]. This geometry has recently stimulated significant interest in the fundamental question whether quantum mechanical tunneling processes between metallic structures can induce these significant changes in the optical response[58–61]. This work provides further support for this hypothesis. More importantly from a practical perspective, it shows how this unique singular optical behavior can be harnessed to realize the smallest possible active photonic device technology.

**Electrical tuning of the optical resonances of a memristive optical antenna**

Next, we discuss the opportunity to tune the resonant response of a finite-sized Ag nanowire antenna. Figure 3A shows an STEM image of a device with an 880 nm section of Ag nanowire extending from the crosspoint. Figure 3B shows EELS maps for the second ($m$ = 2) and third ($m$ = 3) order resonant plasmonic modes of the Ag antenna at energies of 0.88 eV ($\lambda_0$ = 1409 nm) and 1.14 eV ($\lambda_0$ = 1088 nm) respectively. The ability to spatially map the LDOS at different energy loss values enables one to assign and separately study the different FP resonant modes supported by the antenna structure. This shows one of the major benefits over far-field optical light scattering studies where it is very challenging to separate and quantify contributions from different optical modes to the measured signal. From the FP resonance condition it is clear that for these low-order modes the reflection phases can be comparable to the propagation phase. As such, the easily-observable, electrically-induced changes in $\phi_{CR}$ seen in Fig. 2D can have a very large impact on its resonant scattering properties. The results of the electrical switching on this device are included in Fig. S4, but the response was qualitatively similar to the results in Fig.1B.

We can see that both of the observed resonances redshift in spectral position by 96 and 78 nm for the second and third order peaks respectively. The spectral locations of the SP plasmon peaks in the EELS maps were obtained by deconvolving the spectrum using a Voigt profile with three Lorentzian resonance peaks associated with the first three plasmonic resonant modes.[62] These redshifts are of a similar magnitude as those observed in the white light scattering spectra of Fig 1E after dielectric breakdown. The theoretical values predicted by the Fabry-Pérot model using the reflection phases given in Fig. 2D predict a 65 nm and a 20 nm redshift in the second and third order resonances of a silver nanowire antenna 1090 nm in length respectively, i.e. of a similar magnitude as those found in the experiments. This is a remarkable shift considering that no change to the geometry of the device could be observed by STEM imaging during the breakdown event.

The possibility to electrically induce significant changes in the optical response of integrated optical antennas can be applied in a wide range of practical applications in which tuning of light emission and absorption by quantum emitters is desired. These antennas in single or dense arrays that form metamaterials can also be used to manipulate the flow of light in densely packed crosspoint devices. More fundamentally, the experiments suggest a fundamentally new pathway to switch solid-state antenna devices and paves the way for the realization optoelectronic modulators and cavity tuning devices with an unprecedented footprint below 0.01 $\mu m^2$.

## Methods:

**Device fabrication.** Fabrication of the memristive optical antennas was accomplished by a combination of bottom-up and top-down nanofabrication approaches to fabricate the devices on thin nitride membranes (See Fig. S1). First a set of electron transparent substrates was prepared using commercially obtained silicon nitride membranes (15 nm thick, Ted Pella). Such thin membranes are required to maintain a good energy resolution during EELS. Next a set of dielectric pads were processed using EBL to pattern spin-on glass. Then a series of large gold contact pads (50 nm thick) was placed by EBL and DC sputtering (Fig. S1A and Fig. S2A). Silver nanowires grown chemically by the polyol procedure were randomly dispersed on the prepared substrate (Fig. S1B).[63] The nanowires were coated by atomic layer deposition with ~100 cycles of $Al_2O_3$, using trimethyl aluminum and water vapor as precursor gasses. An image of a silver nanowire coated with 120 cycles of ALD is shown in Figure S3, showing that the Al$_2$O$_3$ grows conformally on the silver nanowires. These nanowires were located with the SEM, and smaller gold electrodes were fabricated to connect the silver nanowires to the large contact pads (Fig. S1C, S2A and S2B). In order to apply bias to only one gold/Al$_2$O$_3$/silver junction, it was necessary to connect electrically one of the two gold lines to the metal core of the Al$_2$O$_3$ coated silver nanowire. This was accomplished by the deposition of platinum using a focused ion beam (FIB), where the gallium ion beam used to assist in the platinum deposition damaged the dielectric and lead to a short, which is then coated with the conductive platinum.

**Device characterization.** Light scattering measurements on the silver antennas protruding from the junctions were carried out using a confocal optical microscope coupled to a true-color CCD (Nikon) and a spectrometer (Spectra Physics) with a cooled CCD (Princeton Instruments, Pixis 1024). The silver nanowire devices were illuminated in the bright field objective (Nikon NA0.9, 100x) with white light from a halogen lamp. The light from the silver nanowire antenna was confocally selected and sent to the spectrometer to acquire quantitative spectral information. The spectral data was normalized by subtracting the spectrum backscattered from the air-glass interface from the spectrum of the antenna, then dividing by the lamp spectrum. The spatial resolution of the confocal collection scheme was approximately 1 μm.

**Acknowledgement**
This work at Stanford was supported by a Multi University Research Initiative (MURI 218 FA9550-12-1-0488) from the AFOSR. A.H. acknowledges support from an Air Force Office of Scientific Research, National Defense Science and Engineering Graduate (NDSEG) Fellowship, 32 CFR 168a.

# Figures

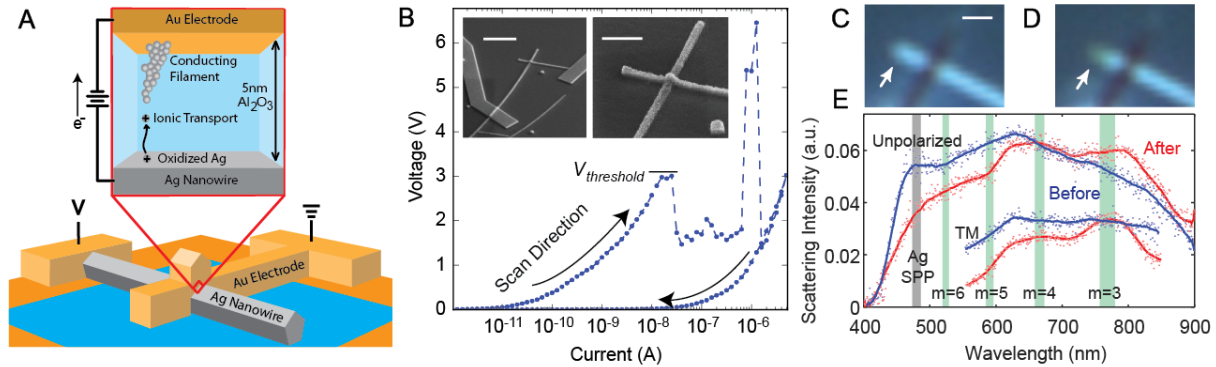

**Figure 1: Electrical and optical switching properties of a memristive optical antenna.** A) Schematic of the device with a metallic crosspoint junction made from a lithographically-defined Au electrode, an $Al_2O_3$ insulating spacer, and a chemically synthesized silver nanowire. The zoom-in illustrates the proposed mechanism of operation involving the formation of a conductive metallic filament in the crossbar junction. B) A typical current-scan on a cross-point junction resulting in dielectric breakdown via the formation of a metallic filament. The formation of the filament results in a state of the device with a significantly lower resistance. Inset: Scanning electron micrographs of the tested crosspoint junction and a zoom-in with scale bars of 2 μm and 600 nm respectively C) Bright-field optical microscopy images showing the change in the scattering response C) before (blue) and D) after (red) electrical switching under unpolarized, white light illumination. The scale bar is 500 nm. E) Confocally-detected white light scattering spectra taken from the crosspoint junction device shown in panels C-D. Both unpolarized and TM polarized spectra are shown. The silver column shows the anticipated spectral location for the localized surface plasmon resonance supported by a subwavelength-diameter silver nanowire. The green columns correspond to the predicted location of different order (m = 3,4,5,6) Fabry-Pérot (FP) resonances of surface plasmon polaritons oscillating between the Ag nanowire truncation and crosspoint. The width of the green beams corresponds to the expected change in resonance frequency upon switching the junction from an FP resonance model.

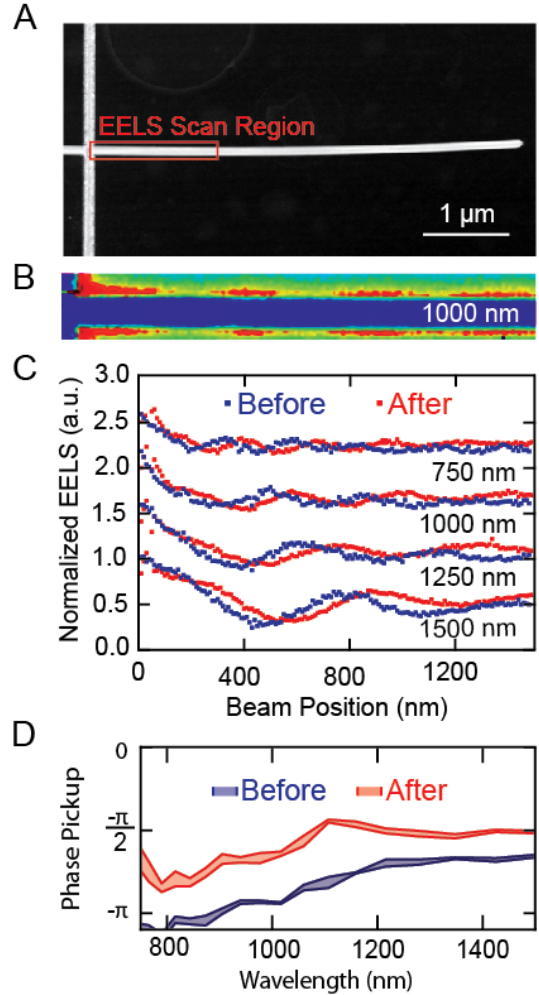

**Figure 2: Analyzing the changes in the local density of optical states near a cross-point junction resulting from an electrical switching event.** A) STEM image of a cross-point junction device from which an Ag nanowire extends well beyond the junction. B) STEM EELs map taken from the area outline in panel A and for an energy corresponding to a free space wavelength of 1000 nm. C) Line profiles of the EELs signal along the Ag nanowire as extracted from STEM maps as shown in panel B. Line profiles are shown at several electron energy losses that correspond to free-space wavelengths in the range from 750 nm to 1500 nm. Line-scans before (blue) and after (red) show the spatial shift in the locations of the antinodes as result of the electrical switching event. D) The changes in the reflection phase of SPPs from the cross-point junction upon switching. The values are extracted from EELs line-scans as shown in panel C. The width of the lines indicates the error in the fits to the experimental EELs data.

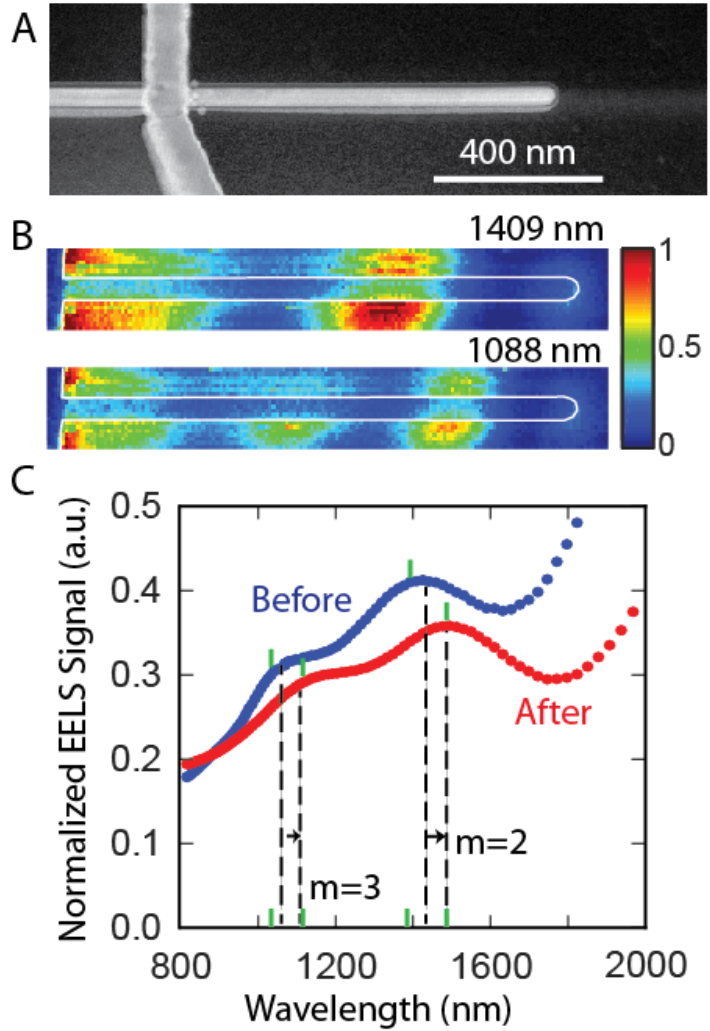

**Figure 3: Electrical tuning of a memristive antenna as verified using EELs maps and EELs spectra.** A) STEM image of a memristive optical antenna after breakdown showing that the overall morphology is intact and that the Au and Ag lines remain morphologically distinct. B) STEM EELs maps for the second and third order resonant modes after switching C) STEM EELs spectra averaged over an area including the antenna showing a redshift in resonant frequencies before and after breakdown. The dashed lines indicate the spectral locations of the Fabry-Pérot resonances for this antenna and the green marks show the deconvolved locations of the EELS plasmon resonances.